\documentclass[conference]{IEEEtran}

\pdfoutput=1
\usepackage{ucs}
\usepackage[utf8x]{inputenc}
\usepackage[english]{babel}
\usepackage{underscore}	  
\usepackage{subfigure}

\usepackage{graphicx}
\DeclareGraphicsExtensions{.pdf,.png,.jpg}

\begin{document}

\title{BitTorrent Swarm Analysis through Automation and Enhanced Logging}

\author{\IEEEauthorblockN{Răzvan Deaconescu, Marius Sandu-Popa, Adriana Drăghici, Nicolae Țăpuș}
\IEEEauthorblockA{Automatic Control and Computers Faculty\\
University Politehnica of Bucharest\\
Bucharest, 060042\\
Email: \{razvan.deaconescu,nicolae.tapus\}@cs.pub.ro, \{marius.sandu-popa,adriana.draghici\}@cti.pub.ro}}

\maketitle

\begin{abstract}
%
%
Peer-to-Peer protocols currently form the most heavily used protocol class in
the Internet, with BitTorrent, the most popular protocol for content
distribution, as its flagship.

A high number of studies and investigations have been undertaken to measure,
analyse and improve the inner workings of the BitTorrent protocol. Approaches
such as tracker message analysis, network probing and packet sniffing have
been deployed to understand and enhance BitTorrent's internal behaviour.

In this paper we present a novel approach that aims to collect, process and
analyse large amounts of local peer information in BitTorrent swarms. We
classify the information as periodic status information able to be monitored
in real time and as verbose logging information to be used for subsequent
analysis. We have designed and implemented a retrieval, storage and
presentation infrastructure that enables easy analysis of BitTorrent protocol
internals. Our approach can be employed both as a comparison tool, as well as a 
measurement system of how network characteristics and  protocol implementation 
influence the overall BitTorrent swarm performance.

We base our approach on a framework that allows easy swarm creation and
control for different BitTorrent clients. With the help of a virtualized
infrastructure and a client-server software layer we are able to create,
command and manage large sized BitTorrent swarms. The framework allows a user
to run, schedule, start, stop clients within a swarm and collect information
regarding their behavior.

\end{abstract}

\textbf{\textit{Keywords -- BitTorrent; swarm analysis; protocol messages;
logging}}



\section{Introduction}
\label{sec:introduction}
%
%


With the exponential growth of digital content and available information,
Peer-to-Peer systems have become the most important protocol class for data
distribution \cite{ipoque}.

Among the wide variety of Peer-to-Peer protocols (Kazaa, DirectConnect,
eDonkey, Kademlia, Gnutella), the BitTorrent protocol has proven to be the
nowadays ``killer protocol''.  With over 30\% of the Internet traffic
\cite{ipoque}, BitTorrent is the most heavily used protocol in the Internet.
The use of simple yet powerful techniques such as tit-for-tat or
rarest-piece-first have selected BitTorrent as the best choice for large data
distribution.

In order to keep up with recent advances in Internet technology, streaming and
content distribution, Peer-to-Peer systems (and BitTorrent) have to adapt and
develop new, attractive and useful features. Extensive measurements, coupled
with carefully crafted scenarios and dissemination are important for
discovering the weak/strong spots in Peer-to-Peer based data distribution and
ensuring efficient transfer.

In this paper we present a framework for running, commanding and managing
BitTorrent swarms. The purpose is to have access to a easy-to-use system for
deploying simple to complex scenarios, make extensive measurements and collect
and analyze swarm information (such as protocol messages, transfer speed,
connected peers) \cite{bt-swarm-analysis}.

\subsection{BitTorrent Keywords}

The heart of the BitTorrent protocol is a \textbf{torrent file}. The torrent
file is a meta-information file containing information regarding the content to
be shared/distributed. Any participant (\textbf{peer}) has to have access to
the torrent file.

An initial peer needs to have access to the complete file for bootstrapping the
transfer. This peer is called the \textbf{initial seeder}. A peer that has
access to the complete content and it's only uploading it is called a
\textbf{seeder}. A peer who is downloading and uploading and has incomplete 
access to the file, is called a \textbf{leecher}.


A collection of peers (seeder or leechers) who are  participating in a transfer 
based on torrent file forms a \textbf{swarm}

The core of the BitTorrent protocol is the \textit{tit for tat} mechanism,
also called \textit{optimistic unchoking} allowing for upload bandwidth to be
exchanged for download bandwidth. A peer is hoping another peer will provide
data, but in case this peer doesn't upload, it will be \textit{choked}.
Another important mechanism for BitTorrent is \textit{rarest piece first}
allowing rapid distribution of content across peers. If a piece of the content
is owned by a small group of peers it will be rapidly requested in order to
increase its availability and, thus, the overall swarm speed and performance.

\subsection{Swarm Management Framework}

\textit{The swarm management framework} is a service-based infrastructure that
allows easy configuration and commanding of BitTorrent clients on a variety of
systems. A client application (\textit{commander}) is used to send
commands/requests to all stations running a particular BitTorrent client. Each
station runs a \textit{dedicated service} that interprets the requests and
manages the local BitTorrent client accordingly.

The framework is designed to be as flexible and expandable as possible. As of
this point it allows running/testing a variety of scenarios and swarms. Based
on the interest of the one designing and running the scenario, one may
configure the BitTorrent client implementation for a particular station, alter
the churn rate by configuring entry/exit times in the swarm, add rate limiting
constraints, alter swarm size, file size etc. Its high reconfigurability allows
one to run relevant scenarios and collect important information to be analyzed
and disseminated.

Through automation and client instrumentation the management framework allows
rapid collection of status and logging information from BitTorrent clients.
The major advantages of the framework are:
\begin{itemize}
  \item \textit{automation} -- user interaction is only required for starting the
clients and investigating their current state;
  \item \textit{complete control} -- the swarm management framework allows the
user/experimenter to specify swarm and client characteristics and to define
the context/environment where the scenario is deployed;
  \item \textit{full client information} -- instrumented clients output detailed
information regarding the inner protocol implementation and transfer
evolution; information are gathered from all client and used for subsequent
analysis.
\end{itemize}

\subsection{Information collection}

Based on the infrastructure we present a novel approach involving client-side
information collection regarding client and protocol implementation. We have
instrumented a libtorrent-rasterbar client~\cite{libtorrent} and a
Tribler~\cite{tribler} client to provide verbose information regarding
BitTorrent protocol implementation.  These results are collected (see
Section~\ref{sec:storage}) and subsequently processed and analysed through a
rendering interface (see Section~\ref{sec:processing}).

Swarm measured data are usually collected from trackers. While this offers a
global view of the swarm it has little information about client-centric
properties such as protocol implementation, neighbour set, number of connected
peers, etc. A more thorough approach has been presented by Iosup et
al.~\cite{mprobe}, using network probes to interrogate various clients.

Our approach, while not as scalable as the above mentioned one, aims to collect
client-centric data, store and analyse it in order to provide information on
the impact of network topology, protocol implementation and peer
characteristics. Our infrastructure provides micro-analysis, rather than
macro-analysis of a given swarm. We focus on detailed peer-centric properties,
rather than less-detailed global, tracker-centric information. The data
provided by controlled instrumented peers in a given swarm is retrieved,
parsed and stored for subsequent analysis. Section~\ref{sec:architecture}
details the modules and information flow in our infrastructure.

We differentiate between two kinds of BitTorrent messages, thoroughly
described in Section~\ref{sec:messages}: \textit{status messages}, which
clients provide periodically to report the current session’s download state,
and \textit{verbose messages} that contain protocol messages exchanged between
peers (chokes, unchokes, peer connections, pieces transfer etc.).

As BitTorrent clients for our experiments, we chose the
libtorrent-rasterbar~\cite{libtorrent} implementation and
Tribler~\cite{tribler}. In our studies~\cite{bt-vi}, libtorrent-rasterbar has
proven to be the fastest BitTorrent client, while Tribler is one of the most
feature rich client from a scientific point of view. Each client outputs
information in a specific format such that a different message parser is
required for each client. Detailed information on the messages and client
instrumentation are presented in Section~\ref{sec:messages}.

Depending on the level of control of the swarm, we define two types of
environments. A \textit{controlled environment}, or \textit{internal swarm}
uses only instrumented controlled clients. We have complete control over the
network infrastructure and peers. A \textit{free environment} or
\textit{external swarm} is usually created outside the infrastructure, and
consists of a larger number of peers, some of which are the instrumented
controlled clients. Our experiments so far have focused on \textit{controlled
environments}; we aim to extend our investigations to \textit{free environment
swarms}.

\subsection{P2P-Next}

This paper is part of the research efforts within the P2P-Next FP7 project
\cite{p2p-next}.


\section{Context}
\label{sec:context}
The proposed swarm management framework was created and designed to provide
data for the \textit{BitTorrent analysis system} presented in
~\cite{bt-swarm-analysis}. This system is focused on offering the means to
collect, store and visualize BitTorrent swarm data at a peer-centric level.
This degree of detail is provided at BitTorrent client level, thus our
experiments aim to gather information about protocol implementation and peer
characteristics.

The framework supports experiments on instrumented BitTorrent clients
(currently only Tribler~\cite{tribler} and Hrktorrent~\cite{hrk} (based on
libtorrent-rasterbar)), which provide the data needed for the analysis system.
These clients run in command-line mode and are configured to output the
communication between peers at a protocol level.

The analysis system consists of parsers and a rendering engine that interact
with a relational database. The messages exchanged between peers and those
output by the client with the state of the transfers, are stored in
\textit{verbose logs} files and \textit{status log} files. The parsers take
these files as input, in order to extract the information provided by each
message and store it into the database. The analysis of protocol messages
coupled with the information regarding the transfer status allows detection of
weak spots of the protocol implementation, thus providing feedback about the
client or possible improvements.

\section{Architecture} \label{sec:arch} 
%
%


\begin{figure}[h]
  \begin{center}
    \includegraphics[width = 3.5in, height = 1.5in]{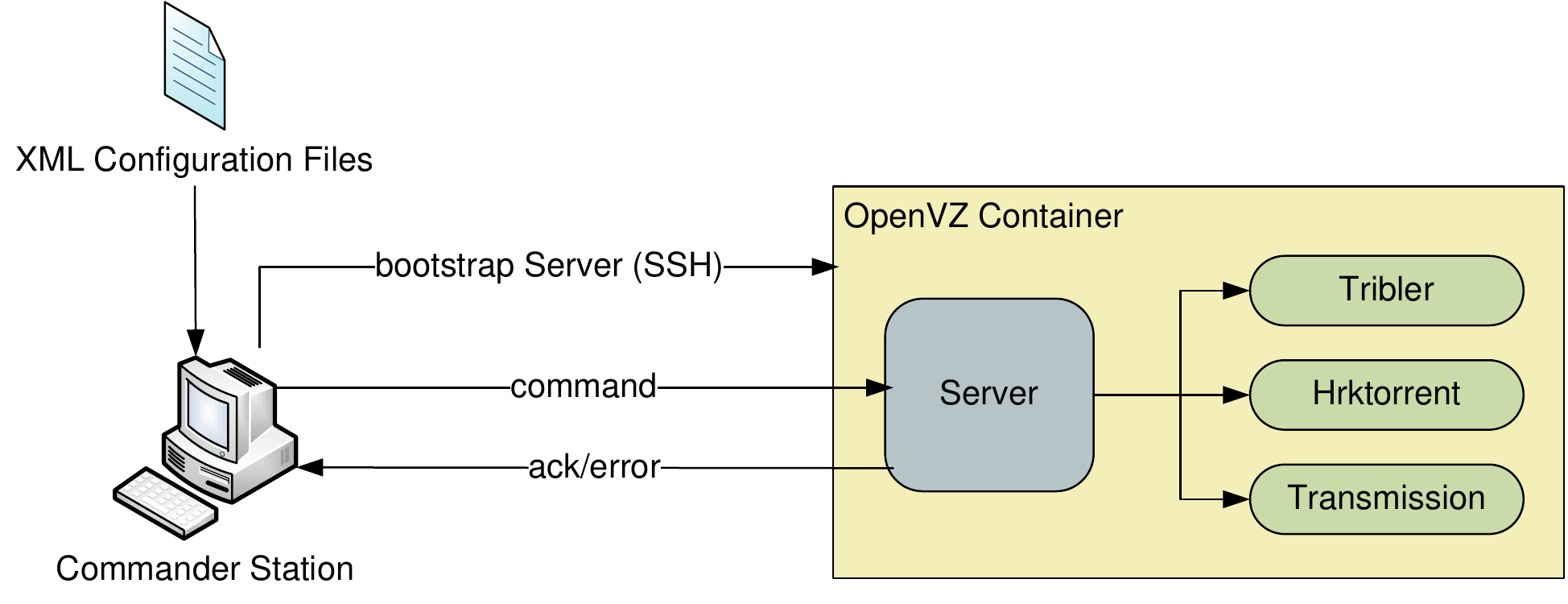}
  \end{center}
  \caption{Software Service System overview}
  \label{fig:schema}
\end{figure}
The software service infrastructure was designed with the goal
of remotely controlling BitTorrent clients. Its architecture(Fig. 1) is built on
a client-server model, with a single client addressed as \textit{Commander} and 
multiple servers. The BitTorrent clients reside in OpenVZ virtual containers and are controlled
only through the \textit{Server} service, by interacting with the Commander interface. A 
SSH connection is used by the Commander for the initial bootstrapping, in case the 
service is not active.

The services are completely implemented in Python, easily allowing extensions
and offering improved maintainability over the shell scripts used in an earlier
virtualized testing environment~\cite{bt-vte}.

The BitTorrent scenarios are defined using XML configuration files which can be
considered as input to the Commander. These files contain information not only
about each container that should be used, but also about the torrent transfers,
like file names and paths. A more through description can be found in
section~\ref{sec:xml}.

In order to examine BitTorrent transfer at a protocol implementation level, we
propose a system for storing and analysing logging data output by BitTorrent
clients. It currently offers support for
hrktorrent/libtorrent~\cite{hrk}~\cite{libtorrent} and Tribler~\cite{tribler}.

Data is provided by BitTorrent clients in log files that are parsed, stored,
intepreted and rendered. We have divided the information generated by clients
into \textbf{status log files} and \textbf{verbose log files}, each composed
of one of two types of messages.

\textit{Status messages} are periodic messages reporting session state.
Messages are usually output by clients at every second with updated
information regarding number of connected peers, current download speed,
upload speed, estimated time of arrival, download percentage, etc. Status
messages are to be used for real time analysis of peer behaviour as they are
lightweight and periodically output (usually every second).

\textit{Verbose messages} or \textit{log messages} provide a thorough
inspection of a client's implementation. The output is usually of large
quantity (hundreds of MB per client for a one-day session). Verbose
information is stored in client side log files and is subsequently parsed and
stored.


Currently, the infrastructure consists of the following modules:

\begin{itemize}
  \item \textbf{Parsers} -- receive log files provided by BitTorrent
clients during file transfers. Due to differences between log file formats,
there are separate pairs of parsers for each client. Each pair analyses status
and verbose messages.
  \item \textbf{Database Access} -- a thin layer between the database system and
other modules.  Provides support for storing messages,  updating and reading
them.
  \item \textbf{SQLite Database} -- contains a database schema with tables
designed for storing protocol messages content and peer information.
  \item \textbf{Rendering Engine} -- consists of a GUI application that
processes the information stored in the database and renders it using plots
and other graphical tools.
\end{itemize}

\begin{figure}[h]
  \begin{center}
    \includegraphics[width = 3.5in, height = 1.5in]{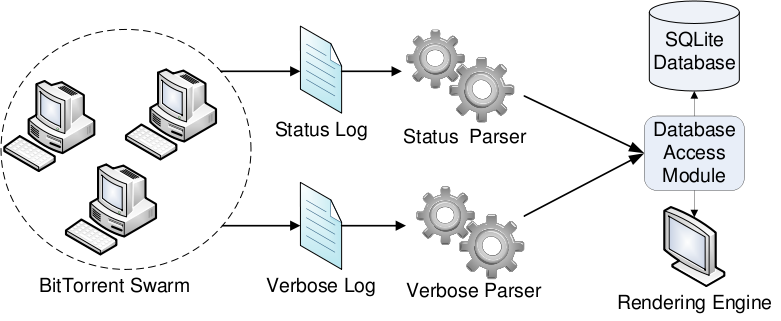}
  \end{center}
  \caption{Logging system overview}
  \label{fig:logarch}
\end{figure}

As shown in figure \ref{fig:logarch}, using parsers specific
to each type of logging file, messages are sent as input to the
\textit{Database Access} module that stores them into an SQLite database. In
order to analyse peer behaviour the Rendering Engine reads stored logging
data using the Database Access module and outputs it to a graphical user
interface. More information on each component is presented in the following
sections.

\subsection{Physical Infrastucture} \label{sec:infra} 

The current setup of the swarm management framework consists of 10 commodity
hardware systems (hardware nodes) each running 10 OpenVZ virtual environments
(VEs), for a total of 100 virtualized systems. Each virtualized system runs a
single Server daemon and a single BitTorrent client.

All hardware nodes are identical with respect to the CPU power, memory
capacity and HDD space and are part of the same network. The network
connections are 1Gbit Ethernet links. Hardware nodes and virtualized
environments are running the same operating system (Debian GNU/Linux
Lenny) and the same software configuration.

To simulate real network bandwidth restrictions we use Linux traffic control
(the \texttt{tc} tool) or client-centric options to limit peer upload/download
speed. As virtualized systems are usually NAT-ed, \texttt{iptables} is also
used on the base stations.

As all stations use common scripts and the same BitTorrent clients, important
parts of the filesystem are accessed through NFS (\textit{Network File
System}). Thus, in case of 100 virtualized systems, only one of them is
actually storing configuration, executable and library files; the other
systems use NFS.

Easy system administration has been ensured through the use of
cluster-oriented tools such as \textit{Cluster SSH} or \textit{Parallel SSH}.

\subsection{XML Configuration Files} \label{sec:xml} 
%
%

As we wanted to make it as easy as possible to deploy new BitTorrent
swarms, we designed our architecture to support two XML configuration files:
one for physical nodes configuration and one for BitTorrent swarms
configuration.

The \textit{nodes} XML file describes the physical infrastructure
configuration. It stores information about:
\begin{itemize}
 \item physical nodes/OpenVZ containers IP addresses and NAT ports \footnote{All the
physical machines in the deployed environment are behind NAT.};
 \item SSH port and username;
 \item Server and Bittorrent clients paths.
\end{itemize}

The \textit{swarm} XML file is used to describe the swarm configuration. It
maps a BitTorrent client to a physical node from the nodes XML configuration
file, and contains the following information:
\begin{itemize}
 \item torrent file for the experiment (same path on all containers)
 \item BitTorrent client upload/download speed limitations.
 \item output options (download path, logs paths)
\end{itemize}

The speed limitations are enforced using the \textit{tc} Linux tool or internal
client bandwidth limitation options.



\subsection{Commander} \label{sec:commander} 
The \textit{Commander} is a command-line tool that provides easy control over 
the BitTorrent clients in our experiments by communicating with the Server daemon. 
It is built entirely in Python and is easily expendable to support new protocol messages
and other features.

The Commander receives as input the two XML configuration files discussed in Section \ref{sec:xml} and 
interacts with the Server through several commands : \textit{bootstrap}, \textit{archive}, 
\textit{start}, \textit{stop}, \textit{status}, \textit{getclients},
\textit{getoutput}, \textit{cleanup}. The \textit{bootstrap} command is made through SSH and 
starts the Server daemon(s).
The other commands use socket communication to a designated port and specific node IP.
Through the Commander, users can send commands to both single or multiple virtualized containers. All commands take
as parameters node and client ids.

\subsection{Server} \label{sec:server} 
The \textit{Server} application represents a daemon~\cite{daemon} that listens
for incoming connections and manages BitTorrent clients. Upon start-up, the
server receives as input from the Commander the IP address on which to bind
itself for socket connections.  The port on which it listens is predefined in a
configuration file visible to both Server and Commander.

Similar to the Commander application, the language chosen for the
implementation is Python, which offers several C-like functionalities, like the
\textit{socket} module for communication and the \textit{subprocess} for
process spawning(the server is responsible for starting and stopping the
BitTorrent clients). The BitTorrent swarm analysis system described in
section~\ref{sec:context} is also entirely implemented in Python, and the
Server uses its status file parsers in order to obtain the latest information
about a transfer status.

The Server is separated from the BitTorrent clients using a thin layer of
classes, implemented for each client, which provide the interface needed for
commanding their execution and establishing their input parameters.


\section{Communication Protocol} \label{sec:proto} 
The system design implies that BitTorrent clients reside on remote machines and
are managed through a \textit{Server} application, which runs as a daemon on
their system. This Server is remotely controlled, being started, restarted and
stopped using SSH commands initiated through the \textit{Commander}
application. Once the Server is started, the Commander acts as its client,
communicating with it in order to control the BitTorrent applications. Our
protocol implies that each BitTorrent client started by the Server is
associated  with only one torrent file.

Currently, the software service infrastructure supports the following messages:
\begin{itemize}
  \item \textit{START-CLIENT} -- the server will start a client with the given
parameters.
  \item \textit{STOP-CLIENT} -- the server will stop a client with the given
identifier.
  \item \textit{GET-CLIENTS} -- the server replies with a list of running
clients.
  \item \textit{GET-OUTPUT} -- the server replies with information about
clients output (running or not)
  \item \textit{ARCHIVE } -- the server creates archives with the files
indicated in the message, and deletes the files.
  \item \textit{GET-STATUS} - returns information about an active transfer.
  \item \textit{CLEANUP} - removes files, extendable to other file types.

The dictionary maps the types of the files that need to be removed, in the
current version of the implementation it supports the following keys:
\begin{itemize}
  \item ALL -- if True, then erases all files related to the experiment
  \item DOWN -- if True, erases all downloaded files
  \item VLOGS -- if True, erases all verbose log files
  \item SLOGS -- if True, erases all status log files
  \item ARCHIVE -- if True, erases all archives related to the experiment
\end{itemize}
\end{itemize}

The Commander initiates transfers by starting a client with a specific torrent
file and options (download path, log files paths and names), and the Server
returns a corresponding ID, which can be used to check the transfer status. The
\textit{status information} is retrieved from the status log files, and
currently supports the following parameters: download speed, upload speed,
downloaded size, uploaded size, eta(estimated time of arrival), number of
peers. In the reply message body, each parameter uses a string identifier
(parameter_name) and is followed by its corresponding value.

\section{Protocol Messages and Client Instrumentation}
\label{sec:messages}
%






The logging system performs in-depth swarm analysis by inspecting protocol
messages exchanged between peers, together with transfer status information
such as upload speed, download speed, download percentage, number of peers.



Our study of logging data takes into consideration two open-source BitTorrent
applications: Tribler~\cite{tribler} and hrktorrent~\cite{hrk} (based
on libtorrent-rasterbar~\cite{libtorrent}). While the latter needed minimal
changes in order to provide the necessary verbose and status data, Tribler had
to be modified significantly.

The process of configuring Tribler for logging output is completely automated
using shell scripts and may be reversed. The source code alterations are
focused on providing both status and verbose messages as client output
information.

\textit{Status message} information provided by Tribler includes transfer
completion percentage, download and upload rates. In the modified version, it
also outputs current date and time, transfer size, estimated time of arrival
(ETA), number of peers, and the name and path of the transferred file.

In order to enable \textit{verbose message} output, we took advantage of the
fact that Tribler uses flags that can trigger printing to standard output for
various implementation details, among which are the actions related to
receiving and sending BitTorrent messages. The files we identified to be
responsible for protocol data are changed using scripts in order to print the
necessary information and to associate it to a timestamp and date. Since most
of the protocol exchange data was passed through several levels in Tribler's
class hierarchy, attention had to be paid to avoid duplicate output and to
reduce file size. In contrast to libtorrent-rasterbar, which, at each transfer,
creates a separate session log file for each peer, Tribler stores verbose
messages in a single file. This  file is passed to the verbose parser, which
extracts relevant parts of the messages and writes them into the database.

Unlike Tribler, hrktorrent's instrumentation did not imply modifying its
source code but defining \texttt{TORRENT\_LOGGING} and
\texttt{TORRENT\_VERBOSE\_LOGGING} macros before building (recompiling)
libtorrent-rasterbar. Minor updates had to be delivered to the compile options
of hrktorrent in order to enable logging output.

The BitTorrent clients and log parsers are configured to distinguish between
the following protocol messages~\cite{bt}:
\begin{itemize}
  \item \textbf{choke} and \textbf{unchoke} -- notification that no data will
be sent until unchoking happens.
  \item \textbf{interested} and \textbf{not interested} -- notifies of a peer's
'interested'/'uninterested' state\footnote{Connections contain two bits of
state on either end: choked or not, and interested or not.}. Data transfer
takes place whenever one side is interested and the other side is not choking.
  \item \textbf{have} -- sent to inform all peers of a piece's  successful
download (its hash matches the one from the .torrent metafile)\footnote{The
peer protocol refers to pieces of the file by index as described in the
metainfo file (.torrent file), starting at zero. Connections contain two bits
of state on either end - choked or not, and interested or not.}.
  \item \textbf{bitfield} -- sent after an initial handshaking sequence between
peers. The payload is a bitfield representing the pieces that have been
successfully downloaded.
  \item \textbf{request} -- sent to obtain blocks of data, the payload
contains a piece index and the block's length and offset within the piece.
  \item \textbf{piece} -- contains a block of data, its position within a piece
and the piece's index. By default these messages are correlated with request
messages, but there are cases when an unexpected piece arrives if choke and
unchoke messages are sent in quick succession and/or transfer is going very
slowly.
  \item \textbf{cancel} - cancels a request for a piece; it has the same
payload as the \textit{request} message. These messages are commonly used when
the download is almost complete; \textit{request} messages are sent to many
peers to make sure the final pieces arrive quickly; when a piece is
downloaded its other requests are cancelled.
\end{itemize}

Although our system processes and stores all protocol message types,
the most important messages for our swarm analysis are those related to
changing a peer's state (choke/unchoke) and requesting/receiving data.
Correlations between these messages are the heart of provisioning information
about the peers' behaviour and BitTorrent clients' performance.

\section{Storage Engine}
\label{sec:storage}
%
%


The swarm analysis infrastructure contains of two levels of storage:
\begin{itemize}
  \item \textit{status and verbose log files} output by clients and sent to
parser modules,
  \item \textit{database storage} populated by parser modules and used by the
rendering interface.
\end{itemize}
Log files are created during a running experiment and parsed after the
experiment had completed. Parsed data is collected as offline
information. All information is subsequently stored in a database file.

The database storage module enables persistence and rapid searching of
relevant information, stored as status data and verbose data. All experiment
data is stored in a single SQLite database file that allows easy migration
and copying.

The storage engine represents an efficient method for collecting information,
compared to using XML files or other file-based approaches. For example, a 5.8
GB worth of text file containing verbose logs was parsed and stored in a
database file of 518 MB.

In addition to holding logging messages, the database stores properties of
BitTorrent clients and details about the swarm (number of peers, number of
initial seeders, start time, file name, file size). It also stores hardware
characteristics about the machine it is running on, such as CPU description,
RAM size, operating system version and network specific information. Along
with these, transfer speed limitations (if any) are stored for each client.

A thin Python layer allows access to the parser and rendering engine for
writing and reading, respectively, data to/from the database. Sample queries
include adding/deleting a new peer, adding/deleting a verbose message (a
BitTorrent protocol~\cite{bt} message), listing messages for a given client in
a specific time frame, listing certain types of BitTorrent messages. In the
current infrastructure, the rendering engine acts as a presentation layer for
collected information.

\section{Result Processing}
\label{sec:processing}
%
%


Once all logging and verbose data from a given experiment is collected, the
next step is the analysis phase. The testing infrastructure provides a GUI
(\textit{Graphical User Interface}) statistics engine for inspecting peer
behaviour. 

The GUI is implemented in Python using two libraries: \textit{matplotlib}
-- for generating graphs and \textit{TraitsUi} -- for handling widgets. It
offers several important plotting options for describing peer behaviour and
peer interaction during the experiment:

\begin{itemize}
  \item \textit{download/upload speed} -- displays the evolution of
download/upload speed for the peer;
  \item \textit{acceleration} -- shows how fast the download/upload speed of
the peer increases/decreases;
  \item \textit{statistics} -- displays the types and amount of verbose
messages the peer exchanged with other peers.
\end{itemize}

The last two options are important as they provide valuable information about
the performance of the BitTorrent client and how this performance is
influenced by protocol messages exchanged by the client.

The \textit{acceleration} option measures how fast a BitTorrent client is able
to download data. High acceleration forms a basic
requirement in live streaming, as it means starting playback of a torrent file
with little delay.

The \textit{statistics} option displays the flow of protocol messages. As
stated in Section~\ref{sec:messages}, we are interested in the choke/unchoke
messages.

The GUI also offers two modes of operation: \textit{``Single Client Mode''},
in which the user can follow the behaviour of a single peer during a given
experiment, and \textit{``Client Comparison Mode''}, allowing for comparisons
between two peers.

\section{Experimental Results}
\label{sec:experiments}
%
%

%
%




\subsection{Experimental Setup}
\label{}

As stated in \ref{sec:context}, the software service infrastructure allows
BitTorrent swarm management. The current implementation was tested on several
scenarios for three clients: Hrktorrent, Tribler and Transmission. The
experiments were conducted on the physical infrastructure presented
in~\ref{sec:infra} and involved checking all the functionalities provided by
the services. The swarms created during these scenarios provided tens of GB of
logging data for the analysis system.

The current setup of our testing infrastructure consists of 10 commodity
hardware systems (hardware nodes) each running 10 OpenVZ virtual environments
(VEs), for a total of 100 virtualized systems. Each virtualized system runs a
single BitTorrent peer.

All hardware nodes are identical with respect to the CPU power, memory
capacity and HDD space and are part of the same network. The network
connections are 1Gbit Ethernet links. Hardware nodes and virtualized
environments are running the same operating system (Debian GNU/Linux
Lenny) and the same software configuration.

To simulate real network bandwidth restrictions we used Linux traffic
control (the \texttt{tc} tool) to limit peer upload/download speed.

\subsection{Results and Measurements}
\label{}

All our experiments have taken place in a \textit{controlled environment} or
\textit{closed swarm}. As such we have had complete control over the peers
and the network topology, allowing us to define the constraints of each
scenario.

We ran several download sessions using the
libtorrent/hrktorrent and Tribler BitTorrent clients and files of different
sizes. All scenarios involved simultaneous downloads for all clients. At the
end of each session, download status information and extensive logging and
debugging information were gathered from each client.

The experiments made use of all 100 virtualized peers which were configured to
use bandwidth limitations. Half of the peers (50) were considered to be
high-bandwidth peers, while the other half were considered to be low-bandwidth
peers. The high-bandwidth peers were limited to 512KB/s download speed and
256KB/s upload speed and the low-bandwidth peers were limited to 64KB/s
download speed and 32KB/s upload speed.


\begin{figure}[h]
  \includegraphics[trim = 3.8cm 4cm 4.3cm 10.7cm, clip, width =
0.45\textwidth]{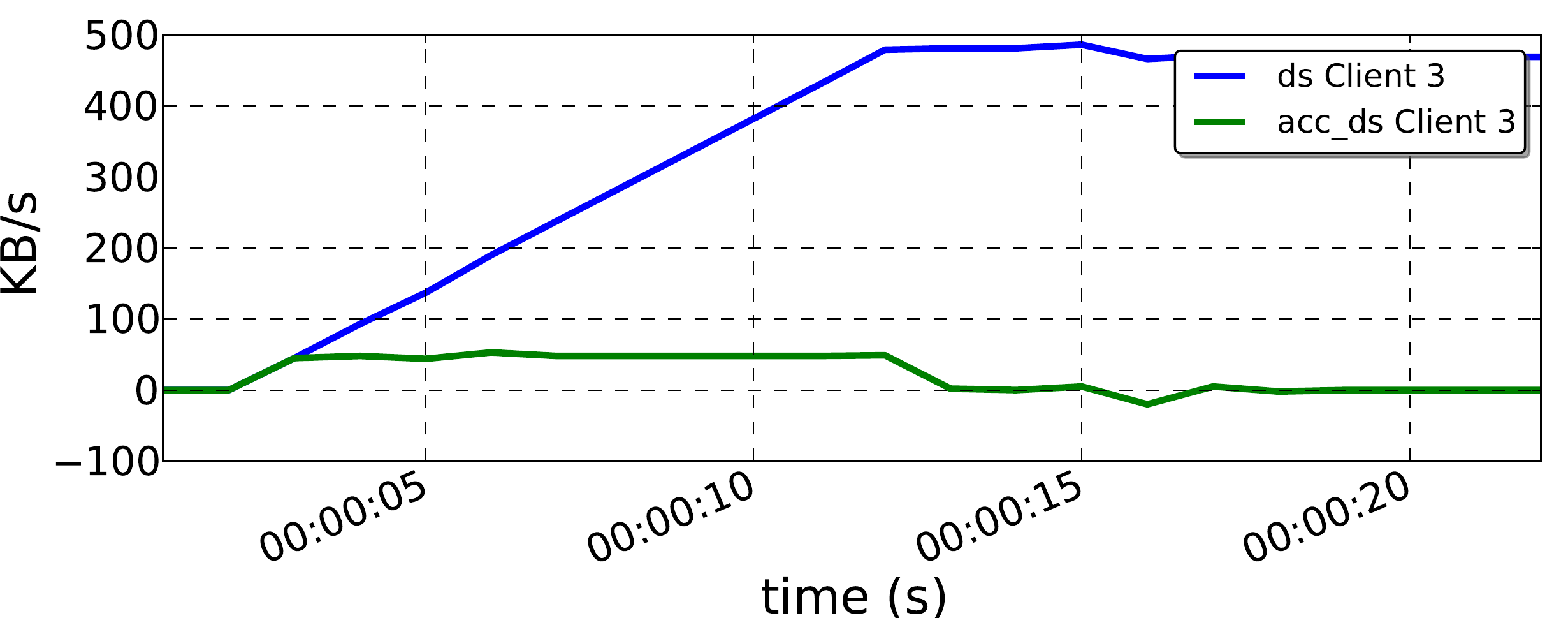}
  \caption{Download speed/acceleration evolution (libtorrent BitTorrent client)}
  \label{fig:down-acc}
\end{figure}

\begin{figure}[h]
    \includegraphics[trim = 15.9cm 13.5cm 4.3cm 3cm, clip, width =
0.45\textwidth]{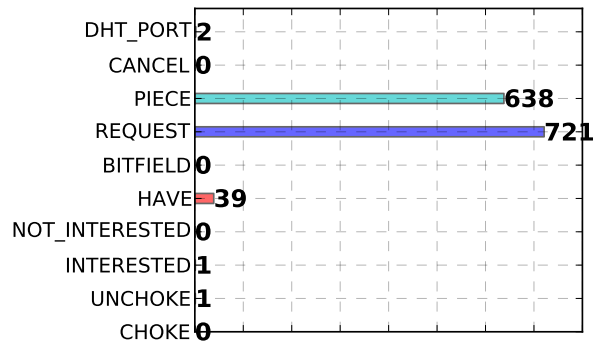}
  \caption{BitTorrent protocol messages (20 seconds)}
  \label{fig:proto-msg}
\end{figure}

Figure~\ref{fig:down-acc} displays a 20 seconds time-based evolution of the
download speed and acceleration of a peer running the libtorrent client.
Acceleration is high during the first 12 seconds, when a peer reaches its
maximum download speed of around 512KB/s. Afterwards, the peer's download
speed is stabilized and its acceleration is close to 0.

All non-seeder peers display a similar start-up pattern. There is an initial
10-12 seconds bootstrap phase with high acceleration and rapid reach of its
download limit, and a stable phase with the acceleration close to 0.

Figure~\ref{fig:proto-msg} displays messages exchanged during the first 20
seconds of a peer's download session, in direct connection with
Figure~\ref{fig:down-acc}. The peer is quite aggressive in its bootstrap phase
and manages to request and receive a high number of pieces. Almost all
requests sent were replied with a block of data from a piece of the file.

The download speed/acceleration time-based evolution graph and the protocol
messages numbering are usually correlated and allow detailed analysis of a
peer's behaviour. Our goal is to use this information to discover weak spots
and areas to be improved in a given implementation or swarm or network
topology.

\section{Related Work}
\label{sec:related}
%
%


As BitTorrent has become the most heavily used peer-to-peer protocol in the
Internet, there have been many measurement studies related to its internals,
enhancements and swarm entities.

Most measurements and evaluations involving the BitTorrent protocol and
applications are either concerned with the behavior of a real-world swarm or
with the internal design of the protocol. There has been little focus on
creating a self-sustained swarm management environment capable of deploying
hundreds of controlled peers, and subsequently gathering results and
interpreting them.

The PlanetLab infrastructure
provides a realistic testbed for
Peer-to-Peer experiments. 
PlanetLab nodes are
connected to the Internet and experiments have a more realistic testbed where
delays, bandwidth and other are subject to change. Tools are also available to
aid in conducting experiments and data collection.

A testing environment involving four major BitTorrent trackers for measuring
topology and path characteristics has been deployed by Iosup et al.
\cite{corr-overlay}. They used nodes in PlanetLab. The measurements were
focused on geo-location and required access to a set of nodes in PlanetLab.

Dragos Ilie et al. \cite{p2p-traf-meas} developed a measurement infrastructure
with the purpose of analyzing P2P traffic. The measurement methodology is based
on using application logging and link-layer packet capture.

One notable study related to BitTorrent protocol analysis
is~\cite{mprobe}. The authors' efforts are directed towards correlating
characteristics of BitTorrent and its Internet underlay, with focus on
topology, connectivity, and path-specific properties. For this purpose they
designed and implemented \textit{Multiprobe}, a framework for large-scale P2P
file sharing measurements. The main difference between their implementation
and our approach is that we focus on an in-depth client-level analysis and not
on the whole swarm.

In~\cite{p2p09} Meulpolder \textit{et al.} present a mathematical model for
bandwidth-inhomogeneous BitTorrent swarms. Based on a detailed analysis of
BitTorrent's unchoke policy for both seeders and leechers, they study the
dynamics of peers with different bandwidths, monitoring their unchoking and
uploading/downloading behavior. Their analysis showed that having only peers
with the same bandwidth is not enough to determine in-depth the peers'
behavior. In those experiments they split the peers into two bandwidth
classes - slow and fast - and they observed that slow ones usually unchoked
other slow peers, their data being transfered from fast peers. Although they
do not offer precise details about the experimental part of monitoring
unchoking behavior and transfers rates, their work relates to what we intend
to do with the logging messages that our system parses and stores. 

While~\cite{p2p09} provides a peer level analysis, another approach is to
study BitTorrent at tracker level, as described in ~\cite{Bardac2009}.  This
paper implements a scalable and extensible BitTorrent tracker monitoring
architecture, currently used in the Ubuntu Torrent Experiment\cite{utorre}
experiment at University Politehnica of Bucharest, the Computer Science and
Engineering Department. The system analyses the peer-to-peer network
considering both the statistic data variation and the geographical
distribution of data. This study is based on a similar infrastructure with the
one we use for our client and protocol level analysis.



\section{Conclusion and Further Work}
\label{sec:conclusion}
%
%


The client-side detailed analysis approach presented earlier is used for
evaluating peer-to-peer swarms and BitTorrent implementations. We have
designed and implemented a message collection and visualisation facility that
allows in-depth analysis of protocol implementations and enhancements.
Several experiments were conducted resulting in large amount of collected data
that were parsed, stored and subjected to analysis through a GUI statistics
engine.

Peer-to-peer measurement infrastructures are commonly using tracker
information or probe-based information, offering an overall view of a swarm.
While not as scalable, our approach allows collection of in depth data
such as low-level protocol information and verbose logging messages. This
requires control of swarm peers, resulting in closed/controlled swarms
providing full information and open/external swarms providing partial
information.

The infrastructure consists of virtualized commodity hardware systems,
instrumented clients that provide extensive information, message parsing
modules, a storage engine and a GUI statistics and interpretation engine. It
allows comparisons between different protocol implementations and studying the
impact of swarm and network characteristics on peer behaviour and overall
swarm performance.

The framework is a service-based infrastructure intended to be used in
conjunction with a result interpretation framework, which collects relevant
information from deployed scenarios and uses that information for analysis and
dissemination.

The advantages of the framework are automation, high degree of control and
access to client logging information regarding protocol internals and transfer
evolution. Realistic scenarios can be deployed and monitored, resulting in
important information provided by client to be subject of subsequent analysis.

As of this point, the framework has been used for internal scenarios. The goal
is to provide the complete infrastructure as a service to be used for running
a wide variety of scenarios. We intend to add scheduling options that allow
users to plan their experiments to be run at a certain time in the future when
enough peers are available.

Traffic shaping is ensured statically at the beginning of each
session. We plan to add a dynamic bandwidth shaping facility that would allow
altering available bandwidth as if there were other communication sessions on
the same link. In order to minimize the administrative configuration, one of
the objectives is to use Linux bridging and connect all virtualized systems
together without the need for NAT.

In order to improve usability, an important objective is to add a web-based
interface to the Commander, which is currently a CLI program. This would
provide the advantage of easy access and configuration of the swarm management
framework.

Currently, the infrastructure supports the hrktorrent~\cite{hrk} and
Tribler~\cite{tribler} implementations. We plan to add support for other
popular open-source clients such as Transmission and Vuze. The open-source
condition is required as client instrumentation is needed to provide in-depth
information.

Extensive simulation and testing and result processing form the major aims
of future planning. We plan to design and run a wide variety of test
scenarios that will result in large amounts of information to be processed
and analysed. Scenarios will focus on measuring the impact of swarm
characteristics on peer behaviour, peer performance and overall swarm
performance. Our current experiments take into account network characteristics
such as bandwidth limitations and swarm characteristics such as client type,
client startup time. We plan to extend these and include the impact of NAT and
firewalled peers, DHT, PEX, peer localisation, network topology, churning,
etc.

As client instrumentation provides in-depth information on client
implementation, it generates extensive input for result analysis. Coupled
with carefully crafted experiments and message filtering, this will allow
the detection of weak spots and of improvement possibilities in current
implementations. Thus it will provide feedback to client and protocol
implementations and swarm ``tuning'' suggestions, which in turn will enable
high performance swarms and rapid content delivery in peer-to-peer systems.

\section*{Acknowledgment}
\label{sec:acknowledgment}

The authors would like to thank Alex Herisanu for kindly providing access to
the NCIT cluster systems we have been using throughout our experiments.

Special thanks go the the P2P-Next~\cite{p2p-next} team who is working
enthusiastically to deliver the next generation peer-to-peer content delivery
platform. Their dedication, professionalism and vision are a constant factor
of motivation and focus for our work.

\bibliographystyle{abbrv}
\bibliography{bt-automation-logging}

\end{document}